\documentclass[twoside]{article}
\usepackage{fleqn,espcrc2}
\usepackage{epsfig}

\usepackage{graphicx}
\usepackage[figuresright]{rotating}

\newcommand{\AmS}{{\protect\the\textfont2
  A\kern-.1667em\lower.5ex\hbox{M}\kern-.125emS}}
\newcommand{\nn}{\nonumber}

\newcommand{\be}{\begin{equation}}
\newcommand{\ee}{\end{equation}}
\newcommand{\bea}{\begin{eqnarray}}
\newcommand{\eea}{\end{eqnarray}}

\newcommand{\evb}{ {\rm eV$^2$} } 
 
\newcommand{\gev}{ {\rm GeV} }

\newcommand{\PCPV}{
\begin{picture}(22,10)
\put(8,-2){\line(2,1){12}}
\put(0,0){$P_{CP}$}
\end{picture}}
\newcommand{\PCPC}{
\begin{picture}(22,10)
\put(0,0){$P_{CP}$}
\end{picture}}
\title{CP-Violation in 3- and 4-family at the Neutrino Factory}

\author{A. Donini\address{
        I.N.F.N., Sezione di Roma, P.le A. Moro 2, I-00185, Rome, Italy}%
        \thanks{This work has been carried out at the 
        Departmento de Fisica Teorica, Universidad Autonoma de Madrid,
        in collaboration with A. Cervera, M.B. Gavela, J.J. Gomez-Cadenas, 
        P. Hernandez, O. Mena and S. Rigolin.} 
       }
       
\begin{document}

\begin{abstract}
The leptonic CP-violating phase $\delta$ can be measured with 
a Neutrino Factory with $2 \times 10^{20} $ useful muons per year, 
if the solar neutrino problem is solved by the LMA-MSW solution, 
with $\Delta m_{12}^2 \ge 2 \times 10^{-5}$ eV$^2$ (in this analysis
a 40 kT magnetized iron detector is considered, taking into account 
its efficiencies and backgrounds). If LSND is confirmed, 
CP-violating phenomena in four-family scenarios can be most easily 
addressed with a small 1 kT detector at $L = \cal O$(10)  km 
(no detailed analysis of the detector efficiencies and backgrounds
has been performed in this case).
\vspace{1pc}
\end{abstract}

\maketitle

\section{Introduction}
Indications in favour of neutrino oscillations have been 
obtained both in solar \cite{solar} and atmospheric \cite{atmospheric} 
neutrino experiments. The atmospheric data require 
$\Delta m_{atm}^2 \sim (2 - 5) \times 10^{-3}$ \evb \cite{neutrino2000}
whereas the solar data prefer $\Delta m_{sun}^2 \sim 10^{-10} - 10^{-4}$ 
\evb, depending on the particular solution to the solar neutrino deficit. 
LSND data \cite{Athanassopoulos:1998pv} would indicate 
a $\nu_\mu \to \nu_e$ oscillation with a third mass difference, 
$\Delta m_{LSND}^2 \sim 1$ \evb. 
Depending on MiniBooNE results \cite{miniboone}, 
the chance to observe a non-zero CP-violating phase in the leptonic sector 
of the Standard Model is completely different. If LSND is not confirmed, 
the experimental results are well described 
by three-family neutrino oscillation. We have a 3 $\times$ 3 mixing
matrix with three angles, $\theta_{12}, \theta_{13}$ and $\theta_{23}$,
and one phase, $\delta$. The CP-violating oscillation probability is
\be
\PCPV = \pm 2 J \left( \sin \Delta_{12} 
+ \sin \Delta_{23} - \sin \Delta_{13} \right) \nn
\ee
with $J = c_{12} s_{12} c^2_{13} s_{13} c_{23} s_{23} \sin \delta$ the Jarlskog
factor and $\Delta_{ij} = \Delta m^2_{ij} L / 2 E_\nu$
(the $\pm$ sign refers to neutrinos/antineutrinos). 
If $\Delta_{12} \ll \Delta_{23}$, $\PCPV$ is negligible.
Therefore, for three-family neutrino mixing the size of the CP-violating
oscillation probability depends on the range of $\Delta m^2_{12}$.
In \cite{Cervera:2000kp,goldenproc} it has been shown that a maximal
phase, $|\delta| = 90^\circ$, can be measured at 90\% CL 
if the LMA-MSW solution with $\Delta m^2_{12} \ge 2 \times 10^{-5}$ \evb
is considered. For smaller values of the solar mass difference, 
it seems impossible to measure $\delta$ with the foreseeable beams. 
If LSND is confirmed, a fourth light sterile neutrino \cite{Pontecorvo:1968fh}
is needed to accommodate the experimental results.
We have in this case a 4 $\times$ 4 mixing matrix with six angles
and three phases. 
In contrast with the three-family case, we can consider CP-violating
observables that do not depend on the solar mass difference. 
Moreover, to maximize $\PCPV$ the baseline is $\cal O$(10) km, 
and the matter effects are negligible. 
In this case, the phases $\delta_i$ could be measured with a small, 
near detector with $\tau$-detection capability 
\cite{Donini:1999jc,3to4proc}.

\section{Three-Family}
\label{3fam}
In ten years from now, we will know $\theta_{23}$ and $\Delta m^2_{23}$
at the 10\% level and at most one of the solutions
to the solar neutrino problem will still be viable (with 
$\theta_{12}$ and $\Delta m^2_{12}$ known within large errors).
However, the present bounds on $\sin^2 2 \theta_{13}$ will go
down to $10^{-2}$, at most. The Neutrino Factory \cite{Albright:2000xi}
is a facility especially
helpful for the measure of the by then (probably) unknown parameters
of the MNS mixing matrix, $\theta_{13}$ and $\delta$. It 
provides high energy and intense $\nu_e (\bar \nu_e)$ beams 
coming from positive (negative) muons which decay 
in the straight sections of a muon storage ring \cite{firstmachine}. 
The $\nu_e (\bar \nu_e)$ can oscillate to $\nu_\mu (\bar \nu_\mu)$,
resulting in ``wrong-sign'' muons in the detector \cite{DeRujula:1999hd}. 
We consider a Neutrino Factory with very intense and energetic muon beam 
($E_\mu = 50$ \gev, $2 \times 10^{20}$ useful muons per year, 
5 years of running with both polarities) and for definiteness
the 40 kT magnetized iron detector described in \cite{cdg};
three reference baselines have been considered, $L = 732, 3500, 7332$ km.
In the LMA-MSW range, the dependence of the oscillation probabilities 
on the solar parameters $\theta_{12}$ and $\Delta m^2_{12}$ and on the 
phase $\delta$ is sizeable at terrestrial distances. 
For this reason it is necessary to perform the simultaneous
measurements of the two unknowns: $\theta_{13}$ and $\delta$.
The matter effects must be properly taken into account: 
if on one side they can be positively used to extract the sign of 
$\Delta m^2_{23}$, on the other hand the fake CP-violating asymmetry 
they induce complicates the measurement of the phase $\delta$.
To reduce the impact of the matter effects, naively a ``short''
($L \le 1000$ km) baseline should be best suited for a CP-violation 
measurement. However, it has been shown in \cite{Cervera:2000kp} 
that an intermediate baseline is more appropriate (see also \cite{others}),
due to the detector efficiency and background (decreasing very fast 
with the distance) and to the correlation between $\theta_{13}$ and $\delta$
at short distance. 
At $L = 3500$ km, with the considered beam-detector setup (and the following
parameter set: $\Delta m^2_{23} = 2.8 \times 10^{-3}$ \evb, 
$\Delta m^2_{12} = 1 \times 10^{-4}$ \evb, $\sin^2 2 \theta_{23} = 1, 
\sin^2 2 \theta_{12} = 0.5$) a precision of a few tenths of degree 
and of a few tens of degrees is attained for $\theta_{13}$ and $\delta$, 
respectively. For $\Delta m^2_{12} \sim 5 \times 10^{-5}$ \evb the phase
$\delta$ can still be measured, but for the lower part of the LMA-MSW
solution, $\Delta m^2_{12} \sim 1 \times 10^{-5}$ \evb the sensitivity
to CP-violation is lost (the detector mass or the intensity of the 
muon beam should be unrealistically increased).
The error induced in the measurement of $\theta_{13}$
by the uncertainty on $\Delta m^2_{12}$ and $\theta_{12}$ can be quite
large. How this could affect the measurement of $\delta$ must be carefully
studied. On the contrary, the Neutrino Factory should reduce 
the uncertainty on $\theta_{23}$ and $\Delta m^2_{23}$ 
down to 1\% through disappearance experiments \cite{Barger:2000jj}. 
This level of uncertainty is not expected to affect significantly our results. 

\begin{figure}
\begin{center}
\epsfig{file=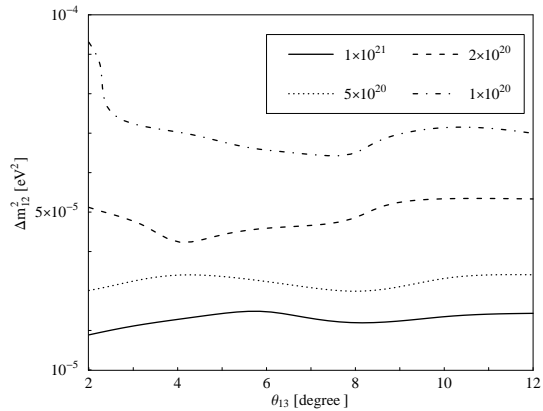,width=75mm}
\end{center} 
\caption{\it Lower limit in $\Delta m_{12}^2$ at which a maximal phase 
($|\delta| = 90^\circ$) can be distinguished from a vanishing phase 
at 99\% CL, as a function of $\theta_{13}$ at $L$ = 3500 km and 
for different numbers of useful muons per year (5 year of running
at both polarities). Background errors and efficiencies are included.}
\label{fig:limdm12}
\end{figure}

We have quantified what is the minimum value of $\Delta m^2_{12}$ for which 
a maximal phase, $|\delta| = 90^\circ$, can be distinguished at 99\% CL 
from $\delta = 0$. The result is shown in Fig.~\ref{fig:limdm12}
for different muon beam intensity. For $2 \times 10^{20}$ useful muons
per year and 5 year of running with both polarities, a maximal phase
can be measured for $\Delta m^2_{12} > 2 \times 10^{-5}$ eV$^2$, 
with very small dependence on $\theta_{13}$ in the range considered
(for $\sin^2 \theta_{13}$ down to $1 \times 10^{-3}$). 

Recently, it has been proposed to measure $\delta$ with a low-energy, 
$E_\nu < 1$ \gev, conventional \cite{lowconv}
or neutrino-factory-like \cite{lownufact} beam and a near detector, 
$L = \cal O $(100) km. This should reduce the fake CP-asymmetry 
induced by matter effects whilst enhancing the CP-violating oscillation 
probability. However, in this case no detailed analysis taking
into account the characteristics of a specific detector has been performed. 
Finally, the reach of a medium-high energy ($E_\nu \ge 1$ \gev)
conventional ``superbeam'' with different detector-types (including
backgrounds) has been studied in \cite{Barger:2000nf}.
Their results show that is very unlikely that $\delta$ can be measured 
with these facilities, except for a small region of parameter space
with $\Delta m^2_{12} \ge 1 \times 10^{-4}$ and 
$\sin^2 \theta_{13} \ge 2 \times 10^{-2}$ where a maximal phase
can be distinguished at 99\% CL from $\delta = 0, \pi$ with 
a JHF-like beam on a supermassive water Cherenkov at $L = 295$ km. 

\section{Four-Family}
\label{4fam}
As in the standard three-family scenario, to observe CP-odd effects 
in oscillations it is necessary to have both physical CP-odd phases 
and at least two non-vanishing mass differences. In contrast with the 
three-neutrino case, the solar suppression 
is now replaced by a less severe atmospheric suppression. CP-violating 
effects are expected to be at least one order of magnitude larger than 
in the standard case, because they are not suppressed by the solar 
mass difference, $\Delta m^2_{12}$.

We consider the same Neutrino Factory as for the three-family case, 
but with only one year of running for both polarities.
For illustration we consider in what follows a generic 1 kT detector 
located at $O(10)$ km distance from the neutrino source with $\tau$-detection
capability \cite{Donini:1999jc}.

In a four-family scenario, the mixing matrix is a 4 $\times$ 4 unitary
matrix with six angles and three phases. In the ``two mass scale dominance'' 
scheme, we neglect the solar mass difference
and end up with a reduced parameter space, consisting of five angles and two 
phases. We consider the 2+2 scenario (two light neutrinos 
with solar mass difference and two heavy neutrinos with atmospheric
mass difference, and $\cal O$(1) \evb LSND separation between the two
pairs), with the ``conservative'' assumption of small cross-gap angles
$\theta_{13},\theta_{14}, \theta_{23}, \theta_{24}$. 
For definiteness, the chosen parameter set is Set 2 of 
\cite{Donini:1999jc,3to4proc}:
$\theta_{34} = 45^\circ$, $\theta_{ij} = 2^\circ$ and 
$\Delta m^2_{atm} = 2.8 \times 10^{-3}$ \evb for 
$\Delta m^2_{LSND} = 1$ \evb.

We define as in \cite{DeRujula:1999hd,Cervera:2000kp} 
the neutrino-energy integrated asymmetry: 
\be
A^{CP}_{\mu \tau} (\delta) = \frac{ \{ N[\tau^+]/N_o[\mu^+] \} 
                                         - \{ N[\tau^-]/N_o[\mu^-] \} }{
                                           \{ N[\tau^+]/N_o[\mu^+] \} 
                                         + \{ N[\tau^-]/N_o[\mu^-] \} }\; ,
\nn
\label{intasy}
\ee
with $N[\tau^\pm]$ the measured number of taus, and 
$N_o[\mu^\pm]$ the expected number of muons 
charged current interactions in the absence of oscillations.
In order to quantify the significance of the signal, we compare the 
value of the integrated asymmetry with its error, in which we include the 
statistical error and a conservative background estimate at
the level of $10^{-5}$. 
The size of the signal-over-noise ratio is very different for $\mu$-
and $\tau$-appearance channels, the difference following the fact that 
the CP-even transition probability $\PCPC (\nu_\mu \to \nu_\tau)$ 
is smaller than $\PCPC (\nu_e \to \nu_\mu)$, due to a stronger suppression 
in small mixing angles.
Notice that the opposite happens in the three-family case, where the 
preferred channel to measure CP-violation is $\nu_e \to \nu_\mu$.

\begin{figure}[t]
\vspace{0.1cm}
\hspace{1cm}
\epsfig{figure=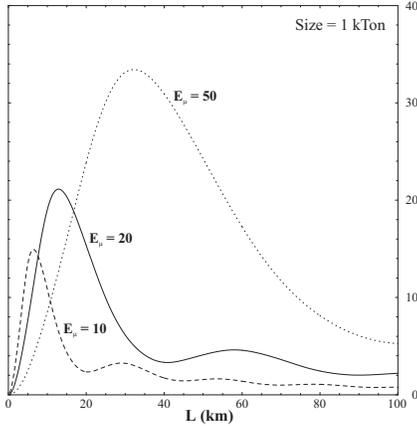,width=5.5cm,angle=0} 
\caption{\it{
Signal over statistical uncertainty for CP violation,
$|A^{CP}_{\mu \tau} (\pi/2) -A^{CP}_{\mu \tau} (0)|/\Delta A^{CP}$,
in the $\nu_\mu \to \nu_\tau$ channel, for the set of parameter in the text. 
We consider a 1 kT detector and $2 \times 10^{20}$ useful muons per year.}} 
\label{CPmtfig12}
\end{figure}

In Fig.~\ref{CPmtfig12} we show the signal-over-noise ratio in $\nu_\mu 
\to \nu_\tau$ versus $\bar\nu_\mu \to \bar\nu_\tau$ oscillations as a 
function of the distance. Matter effects, although negligible, 
have been included, as well as the exact formulae for the probabilities.
For the scenario and distances discussed here, the scaling laws are analogous 
to those derived for three neutrino species in vacuum. 
The maxima of the curves move towards larger distances when the energy of the 
muon beam is increased, or the assumed LSND mass difference is decreased. 
Increasing the energy enhances the significance of the effect at the 
maximum. 
At $E_\mu = 50$ GeV, 30 standard deviation (sd) signals are attainable at 
$L \simeq 40$ km, levelling off at larger distances and finally diminishing 
when $E_\nu/L$ approaches the atmospheric range.

\section{Conclusions}

If LSND is not confirmed, it is possible to measure a maximal
phase $|\delta| = 90^\circ$ at a very intense high-energy ($E_\mu = 50$
\gev ) Neutrino Factory, if the solar neutrino problem is found to be 
solved by the LMA-MSW and $\Delta m^2_{12} \ge 2 \times 10^{-5}$, 
for $\sin^2 \theta_{13}$ as low as $1\times 10^{-3}$. 
These results have been obtained carefully taking into account the 
backgrounds and efficiencies of a magnetized iron detector, finding that 
the optimal baseline for a simultaneous measurement of $\theta_{13}$ 
and $\delta$ is $ L \sim $ 3000 km. Medium-high--energy ($E_\nu \ge 1$ \gev)
conventional ``superbeam'' can possibly measure a maximal phase
if $\sin^2 \theta_{13} \ge 2 \times 10^{-2}$ and 
$\Delta m^2_{12} \ge 1 \times 10^{-4}$ with a
supermassive water Cherenkov detector and a short baseline, $L \sim $ 300 km. 
Low--energy conventional or neutrino-factory--like beams capability
deserve better study with particular attention to the details of 
a specific detector, including background and efficiency. 

If LSND is confirmed by MiniBooNE, a not-so-intense high--energy
Neutrino Factory can most likely measure a maximal phase, regardless
of the solar neutrino problem solution, with a small, $\cal O$(1) kT, 
near detector, $L = \cal O$(10) km, with good $\tau$-identification
capability.

\end{document}